\documentclass[lettersize,journal]{IEEEtran}
\usepackage{amsmath,amsfonts}
\usepackage{algorithmic}
\usepackage{algorithm}
\usepackage{array}
\usepackage[caption=false,font=normalsize,labelfont=sf,textfont=sf]{subfig}
\usepackage{textcomp}
\usepackage{stfloats}
\usepackage{url}
\usepackage{verbatim}
\usepackage{graphicx}
\usepackage{cite}
\begin{document}

\title{Learned Score-Based Turbo Message Passing for Channel Recovery From Spatially Compressed Measurements}

\author{Yiming Wei,~\IEEEmembership{Student Member,~IEEE} and
Xiaojun Yuan,~\IEEEmembership{Fellow,~IEEE}
\thanks{Yiming Wei and Xiaojun Yuan are with the National Key Laboratory of Science and Technology on Communications, University of Electronic Science and Technology of China, Chengdu, China.}}

\markboth{Preprint}%
{Wei and Yuan: Learned Score-Based Turbo Message Passing for Channel Recovery}

\maketitle

\begin{abstract}
Hybrid beamforming (HBF) reduces the hardware cost of massive MIMO base stations, but its channel acquisition requires recovering the full spatial-domain channel from highly compressed beam-domain observations. This paper proposes a learned score-based turbo message passing (L-STMP) framework for uplink HBF channel recovery under extreme spatial compression. By deeply unfolding the turbo message-passing loop and introducing trainable calibration parameters and learnable score priors in each layer, L-STMP liberates STMP from the idealized compression-matrix requirements of right-rotational invariance, and reduces inference complexity to only several unfolded layers. In addition, to preserve dominant channel components that random discrete Fourier transform (DFT) selection may miss, we introduce precoding matrix indicator (PMI)-assisted beam selection as a low-overhead directional prior. Simulation results demonstrate that L-STMP consistently outperforms existing baselines and obtains substantial reconstruction gains under PMI-assisted compression in channel reconstruction.
\end{abstract}


\section{Introduction}
\IEEEPARstart{M}{assive} multiple-antenna technology is a key enabler for improving spectral efficiency, coverage, and spatial multiplexing gains in 5G-Advanced and future 6G systems \cite{heath2015overview}. However, as the number of base-station antennas increases, fully digital architectures that assign an independent radio-frequency (RF) chain, analog-to-digital/digital-to-analog converter (ADC/DAC), and mixer to each antenna become prohibitive in hardware cost, power consumption, and circuit complexity \cite{heath2015overview,alkhateeb2014channel}. Hybrid analog-digital beamforming (HBF) alleviates this burden by using far fewer RF chains than antennas, especially in millimeter-wave, terahertz, and extremely large-scale array systems \cite{alkhateeb2014channel,dai2016beamspace}. This hardware-efficient design, however, forces the base station to observe uplink pilots through a low-dimensional analog or beam-domain projection. Recovering the full spatial-domain channel from such compressed observations is therefore a highly underdetermined inverse problem, whose accuracy directly affects digital precoding, scheduling, interference suppression, and link adaptation.

Existing HBF channel recovery methods mainly exploit angular or beam-domain sparsity and formulate the problem as compressed sensing \cite{alkhateeb2014channel,dai2016beamspace,gao2016channel}. Representative approaches include orthogonal matching pursuit/distributed compressive sensing (OMP/DCS)-based support detection \cite{tropp2007omp,baron2009dcs}, approximate message passing/learned approximate message passing (AMP/LAMP)-based message passing \cite{donoho2009amp,borgerding2017lamp}, and sparse Bayesian learning (SBL)-based Bayesian recovery \cite{wipf2004sbl}. However, under extreme spatial compression where only a small subset of beams is retained, these methods degrade severely: grid-based sparse models suffer from power leakage and beam shift~\cite{gao2016channel,he2020beamspace}, dominant-path supports may be missed by aggressive beam-domain subsampling~\cite{dai2016beamspace,he2018deep}, and even learned priors remain tied to channel sparsity assumptions that inadequately capture real-world channel distributions~\cite{he2018deep,wei2019deep,he2020beamspace}.
Score-based generative models provide a more expressive prior for high-dimensional channel recovery by learning the gradient of the log channel distribution \cite{arvinte2022mimo}. However, diffusion-style posterior sampling typically requires hundreds of reverse-time iterations \cite{kim2026dps}. Score-based turbo message passing (STMP) offers a low-latency alternative~\cite{ma2015turbo,cai2025score}, but its denoising module requires the input to follow an additive white Gaussian noise (AWGN) model, which holds only when the compression matrix is right-rotationally invariant and the system dimension goes to infinity. In practical HBF systems, however, the finite-dimensional compression matrices --- especially structured beam-selection designs --- can hardly meet these conditions, causing the recovery performance to degrade severely.
\IEEEpubidadjcol

The above limitation motivates the learned STMP (L-STMP) receiver proposed in this paper. Firstly, we deeply unfold the turbo message-passing loop into a multi-layer network and introduce trainable calibration parameters and learnable score priors in each layer. This design liberates STMP from the idealized requirements of infinite system dimension and right-rotationally invariant compression matrix, thereby leaving room for flexible analog beamforming design. We show that L-STMP reduces the inference complexity from upto hundreds of iterations of STMP to only several unfolded layers. Secondly, although random discrete Fourier transform (DFT) beam selection approximately satisfies right-rotational invariance, it may miss dominant paths under extreme compression~\cite{dai2016beamspace,3gpp38214}. To mitigate this loss, we introduce precoding matrix indicator (PMI)-assisted beam selection as a low-overhead directional prior, where the PMI indicates the preferred codebook-based precoding direction in the channel state information (CSI) feedback mechanism of the 3rd Generation Partnership Project (3GPP) 5G New Radio (NR) standard~\cite{3gpp38214}. Simulation results show that L-STMP obtains concrete reconstruction gains under this PMI-assisted analog beamforming design, even though the resulting compression matrix clearly violates the property of  right-rotational invariance.

\section{System Model and Problem Formulation}
\label{sec:system_model}

\subsection{Uplink Channel and Compressed Observation}
\label{subsec:compressed_observation_model}

We consider a single-user uplink HBF system where a single-antenna user equipment transmits uplink pilots over $N_{\mathrm{sc}}$ subcarriers, and a base station equipped with an $N_{\mathrm{v}}\times N_{\mathrm{h}}$ uniform planar array of $N_{\mathrm{r}}$ receive antenna elements estimates the full spatial-domain channel from compressed analog observations obtained through a limited number of RF chains. The uplink channel from the user to the base station is denoted by
\begin{equation}
    \mathbf{H}
    =
    \left[
    \mathbf{h}_{1},\mathbf{h}_{2},\ldots,\mathbf{h}_{N_{\mathrm{sc}}}
    \right]
    \in \mathbb{C}^{N_{\mathrm{r}}\times N_{\mathrm{sc}}},
\end{equation}
where $\mathbf{h}_{k}\in\mathbb{C}^{N_{\mathrm{r}}}$ is the receive-array channel vector on the $k$-th subcarrier.


Due to the limited number of RF chains in the HBF receiver, the base station observes the uplink pilot through a low-dimensional spatial projection. Let $M$ denote the number of compressed spatial measurements per subcarrier, where $M \ll N_{\mathrm{r}}$, and let $\mathbf{B}\in\mathbb{C}^{M\times N_{\mathrm{r}}}$ denote a known analog spatial compression matrix. The matrix $\mathbf{B}$ represents the effective RF-domain combining or sampling operation at the receiver. The received signal is
\begin{equation}
    \widetilde{\mathbf{Y}}
    =
    \mathbf{B}
    \mathbf{H}
    \mathbf{X}
    +
    \widetilde{\mathbf{N}},
    \label{eq:observation_with_pilot}
\end{equation}
where $\mathbf{X}=\mathrm{diag}\left(x_{1},\ldots,x_{N_{\mathrm{sc}}}\right)$ collects the uplink pilot symbols transmitted on the $N_{\mathrm{sc}}$ subcarriers and $\widetilde{\mathbf{N}}$ is an AWGN matrix with the power of each element being 
$\sigma^{2}$. The pilot symbols are known at the receiver and have constant modulus with unit power, i.e., $|x_{k}|=1,\forall k$. Normalizing the received signal by right-multiplying $\mathbf{X}^{H}$ removes the pilot term and yields
\begin{equation}
    \mathbf{Y}
    =
    \mathbf{B}
    \mathbf{H}
    +
    \mathbf{N},
    \label{eq:compressed_observation}
\end{equation}
where $\mathbf{Y}\in\mathbb{C}^{M\times N_{\mathrm{sc}}}$, and $\mathbf{N}=\widetilde{\mathbf{N}}\mathbf{X}^{H}$ remains i.i.d. complex Gaussian noise with the same variance $\sigma^{2}$ as $\widetilde{\mathbf{N}}$, by noting that $\mathbf{X}$ is unitary.

\subsection{Problem Formulation}
\label{subsec:problem_formulation}

Given the pilot-normalized compressed observation $\mathbf{Y}$ and the known compression matrix $\mathbf{B}$, the objective is to recover the full complex spatial-domain channel matrix $\mathbf{H}$. The proposed estimator is written as
\begin{equation}
    \widehat{\mathbf{H}}
    =
    f_{\Theta}
    \left(
    \mathbf{Y},
    \mathbf{B}
    \right),
    \label{eq:estimator}
\end{equation}
where $f_{\Theta}(\cdot)$ denotes the learned network with trainable parameters $\Theta$. The spatial observation ratio is $\rho = M / N_{\mathrm{r}}$. Since $M \ll N_{\mathrm{r}}$ and $\rho \ll 1$ in the extreme compression regime of interest, this inverse problem is severely underdetermined and requires both measurement consistency and effective channel priors.

\section{Learned Score-Based Turbo Message Passing}
\label{sec:learned_stmp}

\subsection{Network Overview}
\label{subsec:stmp_overview}

\begin{figure*}[!t]
        \centering
        \includegraphics[width=\textwidth]{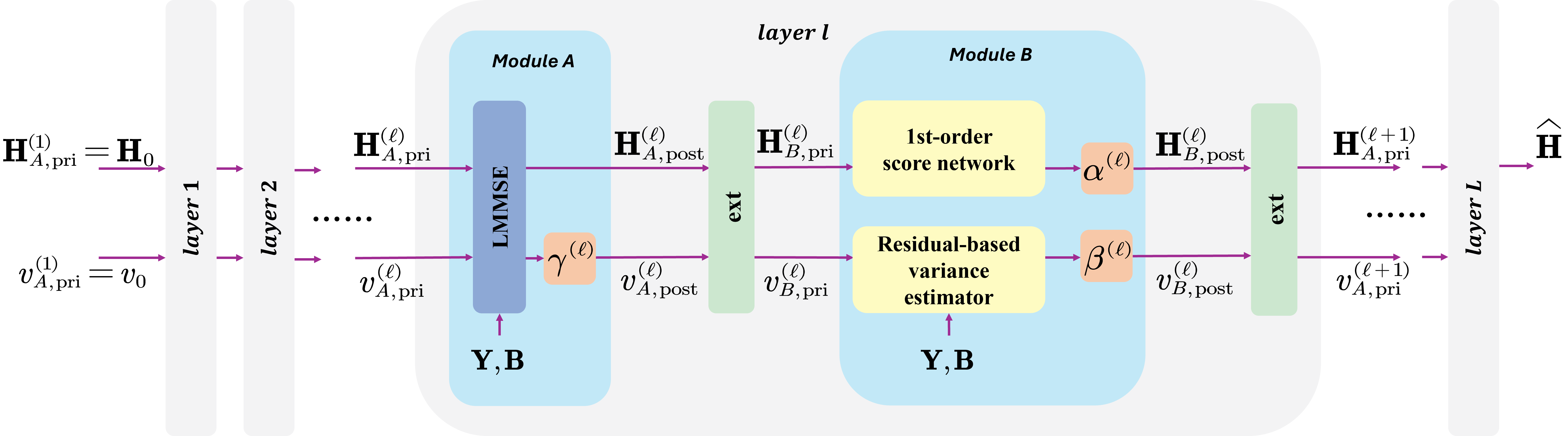}
        \caption{Learned score-based turbo message passing network.}
        \label{fig:learned_stmp}
        \end{figure*}

STMP~\cite{cai2025score} alternates measurement-consistency and score-denoising updates, but its score-based denoiser relies on an AWGN assumption that holds only when the compression matrix is right-rotationally invariant and the system dimension goes to infinity. To relax these requirements, we unfold STMP into a finite-depth trainable network, whose calibration factors and score networks are learned. As shown in Fig.~\ref{fig:learned_stmp}, each layer follows the two-module structure of STMP: Module A enforces consistency with the compressed observation $\mathbf{Y}$ and the known matrix $\mathbf{B}$, while Module B exploits learned score-based channel priors. The compression matrix $\mathbf{B}$ is fixed by the receiver architecture and is not trained. The trainable components are the unfolded-layer calibration parameters and the score-prior networks.

The unfolded network contains $L$ layers and operates directly on the complex-domain observation model. At the $\ell$-th layer, the input to Module A is denoted by $\left(\mathbf{H}_{A,\mathrm{pri}}^{(\ell)},v_{A,\mathrm{pri}}^{(\ell)}\right)$, where $\mathbf{H}_{A,\mathrm{pri}}^{(\ell)}$ is the prior channel estimate and $v_{A,\mathrm{pri}}^{(\ell)}$ is the associated variance. For compactness, variance variables are written as scalars; they can also represent channel-wise variance parameters that are broadcast to the channel matrix dimensions. The first-layer prior is initialized as $\left(\mathbf{H}_{A,\mathrm{pri}}^{(1)},v_{A,\mathrm{pri}}^{(1)}\right)=\left(\mathbf{H}_{0},v_{0}\right)$. 


\subsection{Module A: Observation-Consistency Module}
\label{subsec:observation_consistency}

Given $\mathbf{Y}$ and $\mathbf{H}_{A,\mathrm{pri}}^{(\ell)}$, Module A first computes the residual
\begin{equation}
    \mathbf{R}^{(\ell)}
    =
    \mathbf{Y}
    -
    \mathbf{B}
    \mathbf{H}_{A,\mathrm{pri}}^{(\ell)}.
    \label{eq:stmp_residual}
\end{equation}
Module A then updates the posterior mean by back-projecting the residual through the adjoint of $\mathbf{B}$. When the compression matrix satisfies the row-orthogonality condition $\mathbf{B}\mathbf{B}^{H} = \mathbf{I}_{M}$, the linear MMSE estimate of $\mathbf{H}_{A,\mathrm{post}}^{(\ell)}$ is given by
\begin{equation}
    \mathbf{H}_{A,\mathrm{post}}^{(\ell)}
    =
    \mathbf{H}_{A,\mathrm{pri}}^{(\ell)}
    +
    \frac{
    v_{A,\mathrm{pri}}^{(\ell)}
    }{
    v_{A,\mathrm{pri}}^{(\ell)}
    +
    \sigma^{2}
    }
    \mathbf{B}^{H}
    \mathbf{R}^{(\ell)},
    \label{eq:stmp_a_post_mean}
\end{equation}
and the analytical posterior variance is given by
\begin{equation}
    v_{A,\mathrm{post,ana}}^{(\ell)}
    =
    v_{A,\mathrm{pri}}^{(\ell)}
    -
    \rho
    \frac{
    \left(
    v_{A,\mathrm{pri}}^{(\ell)}
    \right)^{2}
    }{
    v_{A,\mathrm{pri}}^{(\ell)}
    +
    \sigma^{2}
    }.
    \label{eq:stmp_a_post_var_ana}
\end{equation}
Consistent with the $\gamma^{(\ell)}$ block in Fig.~\ref{fig:learned_stmp}, a positive trainable calibration factor is introduced to compensate for variance mismatch under extreme spatial compression:
\begin{equation}
    v_{A,\mathrm{post}}^{(\ell)}
    =
    \gamma^{(\ell)}
    v_{A,\mathrm{post,ana}}^{(\ell)}.
    \label{eq:stmp_a_post_var}
\end{equation}
The green ``ext'' block after Module A removes the incoming prior information and generates the prior of Module B:
\begin{equation}
    v_{B,\mathrm{pri}}^{(\ell)}
    =
    \left(
    \frac{1}{v_{A,\mathrm{post}}^{(\ell)}}
    -
    \frac{1}{v_{A,\mathrm{pri}}^{(\ell)}}
    \right)^{-1},
    \label{eq:stmp_b_pri_var}
\end{equation}
and
\begin{equation}
    \mathbf{H}_{B,\mathrm{pri}}^{(\ell)}
    =
    v_{B,\mathrm{pri}}^{(\ell)}
    \left(
    \frac{
    \mathbf{H}_{A,\mathrm{post}}^{(\ell)}
    }{
    v_{A,\mathrm{post}}^{(\ell)}
    }
    -
    \frac{
    \mathbf{H}_{A,\mathrm{pri}}^{(\ell)}
    }{
    v_{A,\mathrm{pri}}^{(\ell)}
    }
    \right).
    \label{eq:stmp_b_pri_mean}
\end{equation}

\subsection{Module B: Score-Prior Module}
\label{subsec:score_prior}

The score-prior module treats $\mathbf{H}_{B,\mathrm{pri}}^{(\ell)}$ as a noisy pseudo-observation of the channel and performs score-based denoising. Let $s_{\theta_{\ell}}(\cdot,\cdot)$ denote the score network in the $\ell$-th layer, where $\theta_{\ell}$ denotes the trainable network parameters. Following the learned STMP formulation in \cite{cai2025score}, the score network takes the pseudo-observation and the prior variance $v_{B,\mathrm{pri}}^{(\ell)}$ as input. As illustrated by Module B in Fig.~\ref{fig:learned_stmp}, the score network refines the channel mean through a Tweedie-form correction:
\begin{equation}
    \mathbf{H}_{B,\mathrm{post}}^{(\ell)}
    =
    \mathbf{H}_{B,\mathrm{pri}}^{(\ell)}
    +
    \alpha^{(\ell)}
    v_{B,\mathrm{pri}}^{(\ell)}
    s_{\theta_{\ell}}
    \left(
    \mathbf{H}_{B,\mathrm{pri}}^{(\ell)},
    v_{B,\mathrm{pri}}^{(\ell)}
    \right),
    \label{eq:stmp_b_post_mean}
\end{equation}
where $\alpha^{(\ell)}$ is a positive trainable scalar that controls the strength of the score update. Instead of propagating the variance through the score network as in STMP~\cite{cai2025score}, the posterior variance is estimated empirically from the residual between the compressed observation and the updated channel estimate:
\begin{equation}
    v_{B,\mathrm{post}}^{(\ell)}
    =
    \beta^{(\ell)}
    \frac{
    \left\|
    \mathbf{Y}
    -
    \mathbf{B}
    \mathbf{H}_{B,\mathrm{post}}^{(\ell)}
    \right\|_{F}^{2}
    -
    M
    N_{\mathrm{sc}}
    \sigma^{2}
    }{
    N_{\mathrm{sc}}
    \left\|
    \mathbf{B}
    \right\|_{F}^{2}
    },
    \label{eq:stmp_b_post_var}
\end{equation}
where $\beta^{(\ell)}$ is a positive trainable scalar that calibrates the residual-based variance estimate.


The extrinsic block after Module B computes the extrinsic information of Module B, using the message-combining formulas similar to \eqref{eq:stmp_b_pri_var} and \eqref{eq:stmp_b_pri_mean} in Module A, to generate the prior $\left(\mathbf{H}_{A,\mathrm{pri}}^{(\ell+1)},v_{A,\mathrm{pri}}^{(\ell+1)}\right)$ of the next layer.
After $L$ layers, the final reconstruction of the channel $\widehat{\mathbf{H}}$ is given by
\begin{equation}
    \widehat{\mathbf{H}}
    =
    \mathbf{H}_{B,\mathrm{post}}^{(L)},
    \label{eq:stmp_output}
\end{equation}

\subsection{Training Objective}
\label{subsec:stmp_training}

The unfolded STMP network is trained progressively layer by layer, with each new layer trained jointly with the preceding layers. The training loss is the weighted normalized mean-square error over the outputs of all available layers, $\mathcal{L}=\sum_{\ell}\lambda_{\ell}\,\mathbb{E}\left[\left\|\mathbf{H}_{B,\mathrm{post}}^{(\ell)}-\mathbf{H}\right\|_{F}^{2}/\left\|\mathbf{H}\right\|_{F}^{2}\right]$, where $\lambda_{\ell}\geq 0$ weights the $\ell$-th layer, with the newest layer as the primary term and the earlier layers providing auxiliary supervision. The optimized parameters include the layer-wise calibration factors $\{\gamma^{(\ell)}, \alpha^{(\ell)}, \beta^{(\ell)}\}$ in each layer and the weights $\theta_{\ell}$ of the score networks, which are instantiated as NCSN++~\cite{song2021score}, pretrained to learn the score of the channel distribution, and then fine-tuned jointly with the calibration factors. The compression matrix $\mathbf{B}$, its adjoint, $\rho$, and the noise variance are fixed by the measurement model. This objective matches the channel-recovery goal in \eqref{eq:estimator}, while the trainable parameters compensate for the deviation of practical HBF compression matrices from the AWGN extrinsic input and right-rotational invariance conditions assumed in the STMP derivation.

\section{Compression Matrix Selection}
\label{sec:compression_matrix_selection}

We now consider the design of the compression matrix $\mathbf{B}$, Following~\cite{3gpp38214}, we employ the 2D-DFT for the choices of beams. Thus $\mathbf{B}$ can be expressed as:
\begin{equation}
    \mathbf{B}
    =
    \mathbf{S}_{\Omega}
    \mathbf{F},
    \label{eq:dft_compression_matrix}
\end{equation}
where $\mathbf{F}=\mathbf{F}_{\mathrm{v}}\otimes\mathbf{F}_{\mathrm{h}}\in\mathbb{C}^{N_{\mathrm{r}}\times N_{\mathrm{r}}}$ is the unitary two-dimensional (2D) DFT matrix, with $\mathbf{F}_{\mathrm{v}}\in\mathbb{C}^{N_{\mathrm{v}}\times N_{\mathrm{v}}}$ and $\mathbf{F}_{\mathrm{h}}\in\mathbb{C}^{N_{\mathrm{h}}\times N_{\mathrm{h}}}$ the normalized DFT matrices along the vertical and horizontal array dimensions, respectively, and $\mathbf{S}_{\Omega}\in\{0,1\}^{M\times N_{\mathrm{r}}}$ selects the rows indexed by $\Omega$ with $|\Omega|=M$. Since $\mathbf{S}_{\Omega}\mathbf{S}_{\Omega}^{H}=\mathbf{I}_{M}$, $\mathbf{B}$ in \eqref{eq:dft_compression_matrix} satisfies $\mathbf{B}\mathbf{B}^{H}=\mathbf{I}_{M}$.

\subsection{Random DFT Beam Selection}
\label{subsec:random_dft_selection}

The random selector samples $\Omega$ uniformly without replacement:
\begin{equation}
    \Omega
    \sim
    \mathrm{Unif}
    \left(
    \left\{
    \Omega\subseteq\{1,\ldots,N_{\mathrm{r}}\}:|\Omega|=M
    \right\}
    \right).
    \label{eq:random_beam_selection}
\end{equation}
Uniform beam selection makes $\mathbf{B}$ approximately right-rotationally invariant, which is favorable for STMP. However, random selection may miss dominant beam-domain components when $M$ is small. To mitigate this loss, we exploit the PMI from standard CSI feedback as a low-overhead directional prior, yielding the PMI-assisted selection described next.

\subsection{PMI-Assisted DFT Beam Selection}
\label{subsec:pmi_dft_selection}

The PMI-assisted selector uses the PMI as a directional prior. Assuming uplink-downlink channel reciprocity as in time-division duplex (TDD) operation, the precoding direction reported by the PMI also indicates the dominant direction of the uplink channel. Following the Type~I single-panel codebook~\cite{3gpp38214}, the reported PMI $q$ indicates a preferred 2D-DFT beam of the receive array, represented by the antenna-domain direction vector $\mathbf{c}_{q}\in\mathbb{C}^{N_{\mathrm{r}}}$. The importance score of the $n$-th DFT beam is the energy of the $n$-th entry of the DFT-domain projection $\mathbf{F}\mathbf{c}_{q}$:
\begin{equation}
    e_{n}
    =
    \left|
    \left[
    \mathbf{F}
    \mathbf{c}_{q}
    \right]_{n}
    \right|^{2},
    \quad
    n=1,\ldots,N_{\mathrm{r}}.
    \label{eq:pmi_beam_energy}
\end{equation}
The PMI-assisted selector then chooses the $M$ beams with the largest scores:
\begin{equation}
    \Omega
    =
    \mathrm{TopM}
    \left(
    \left\{
    e_{n}
    \right\}_{n=1}^{N_{\mathrm{r}}}
    \right).
    \label{eq:pmi_topm_selection}
\end{equation}

Compared with random selection, PMI-assisted Top-$M$ selection uses the same measurement budget but biases the retained beams toward the dominant spatial direction indicated by the PMI. However, because $\Omega$ in~\cite{3gpp38214} is deterministically determined by the PMI rather than drawn uniformly at random, $\mathbf{B}$ no longer satisfies right-rotational invariance, and STMP is expected to degrade under this design.


\section{Numerical Results}
\label{sec:numerical_results}

\subsection{Simulation Setup}
\label{subsec:simulation_setup}

We evaluate the proposed learned STMP (L-STMP) receiver on an uplink channel dataset generated under a 3GPP urban macro (UMa) non-line-of-sight (NLoS) scenario. The channel contains $N_{\mathrm{sc}}=48$ subcarriers and $N_{\mathrm{r}}=256$ receive antennas of a dual-polarized planar array with $N_{\mathrm{v}}\times N_{\mathrm{h}}=16\times8$ antennas per polarization. In the reported simulations, the general compression matrix $\mathbf{B}$ is primarily instantiated as the random DFT beam selection matrix described in Section~\ref{subsec:random_dft_selection}, while the PMI-assisted DFT selection matrix (Section~\ref{subsec:pmi_dft_selection}) is also evaluated for comparison. We consider $M=64$ and $M=32$ spatial measurements per subcarrier, corresponding to four-fold and eight-fold spatial compression, respectively.

The L-STMP network contains $L$ unfolded layers, each with an independent NCSN++ score network~\cite{song2021score} configured under the variance-exploding SDE, with $L=5$ in the reported experiments. In addition to the layer-wise calibration factors $\{\gamma^{(\ell)}, \alpha^{(\ell)}, \beta^{(\ell)}\}$ described in Section~\ref{sec:learned_stmp}, the network is trained layer-by-layer. The layer-wise loss weights are $\lambda_1=\lambda_2=\lambda_3=\lambda_4=0.2$ with $\lambda_5=1.0$.

The transmitter-side SNR is $\mathrm{SNR} = 10\log_{10}(1/\sigma^{2})$ dB,
where $\sigma^{2}$ is the noise variance in \eqref{eq:compressed_observation}. The channel is normalized in simulation, i.e., $\frac{1}{N_{\mathrm{r}}N_{\mathrm{sc}}}\left\|\mathbf{H}\right\|_{F}^{2}=1$, and the pilot symbols have unit power, so that the average transmit power per spatial measurement is unity.

The reconstruction accuracy is measured by the normalized mean-square error (NMSE)
\begin{equation}
    \mathrm{NMSE}
    =
    10\log_{10}
    \left(
    \frac{
    \left\|
    \widehat{\mathbf{H}}
    -
    \mathbf{H}
    \right\|_{F}^{2}
    }{
    \left\|
    \mathbf{H}
    \right\|_{F}^{2}
    }
    \right)
    \ \mathrm{dB}.
    \label{eq:nmse_metric}
\end{equation}

\begin{figure}[!t]
    \centering
    \includegraphics[width=\columnwidth]{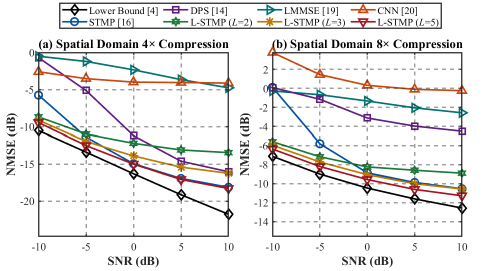}
    \caption{NMSE performance comparison under random DFT beam selection: (a) four-fold spatial compression ($M=64$); (b) eight-fold spatial compression ($M=32$).}
    \label{fig:result_foureight}
\end{figure}

\begin{figure}[!t]
    \centering
    \includegraphics[width=1.0\columnwidth]{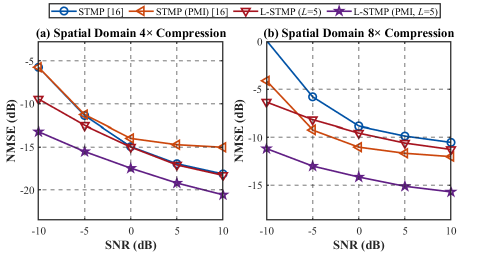}
    \caption{NMSE performance comparison under PMI-assisted DFT beam selection: (a) four-fold spatial compression ($M=64$); (b) eight-fold spatial compression ($M=32$).}
    \label{fig:pmi}
\end{figure}

\subsection{Result Analysis}
\label{subsec:result_analysis}

Fig.~\ref{fig:result_foureight} reports the NMSE performance of the proposed L-STMP receiver and several baselines under random DFT beam selection with four-fold ($M=64$) and eight-fold ($M=32$) spatial compression. The oracle lower bound is obtained by assuming perfect knowledge of the power angle and delay profile of the channel, providing a theoretical performance lower bound~\cite{gao2016channel}. The LMMSE estimator~\cite{edfors1996ofdm} recovers the channel by a direct linear projection of the compressed observation. The CNN-based estimator~\cite{dong2019deepcnnchannelestimation} learns a residual convolutional mapping from the adjoint back-projection $\mathbf{B}^{H}\mathbf{Y}$ to the full channel, representing purely data-driven priors. DPS~\cite{kim2026dps} performs diffusion posterior sampling with the same pretrained score prior but requires thousands of reverse steps, and STMP~\cite{cai2025score} runs the classical iterative turbo loop until convergence. Across all evaluated SNR levels, L-STMP consistently achieves the lowest NMSE among practical methods, approaching the oracle lower bound. The performance gap between L-STMP and the STMP baseline widens under eight-fold compression, indicating that the trainable layer-wise calibration factors and learned score priors effectively compensate for the mismatch between ideal STMP assumptions and practical highly compressed observations. LMMSE and CNN suffer severe degradation as the compression ratio increases, confirming that model-agnostic or purely data-driven priors are insufficient under extreme spatial compression. DPS exhibits moderate performance but requires substantially more inference iterations than the unfolded L-STMP network.

Fig.~\ref{fig:pmi} examines the performance under PMI-assisted DFT selection with four-fold and eight-fold compression. Although PMI-assisted selection preserves dominant-path information that random DFT may miss, it deterministically biases the retained beams toward the PMI direction and therefore breaks the right-rotational invariance of the compression matrix assumed in STMP~\cite{cai2025score}. As a result, the STMP baseline gains little from the PMI prior --- its performance is even slightly degraded at four-fold compression compared with the random-DFT case. L-STMP, in contrast, obtains concrete NMSE gains under PMI-assisted compression at both ratios. This is because L-STMP does not rely on the idealized AWGN extrinsic assumption: it uses the actual $\mathbf{B}$ in the measurement-consistency module and learns the calibration factors $\{\gamma^{(\ell)}, \alpha^{(\ell)}, \beta^{(\ell)}\}$ from the data, so the information gain brought by PMI-guided beam selection is efficiently exploited instead of being offset by assumption violations.

Table~\ref{tab:complexity} compares per-sample inference complexity at the reference configuration ($N_{\mathrm{sc}}=48$, $N_{\mathrm{r}}=256$, $M=64$). The dominant cost is the NCSN++ score-network forward pass (${\sim}1.6\times10^{11}$ FLOPs), while the measurement-consistency $\mathbf{B}/\mathbf{B}^{H}$ overhead is negligible (${\sim}8.6\times10^{5}$ FLOPs). L-STMP requires only $L=5$ score forwards, compared with $200$ for STMP and $2000$ for DPS --- approximately $40\times$ and $400\times$ reductions --- yet achieves much superior estimation accuracy. LMMSE and CNN do not involve score networks and require lower FLOPs, but their NMSE degrades severely, as seen from Fig.~\ref{fig:result_foureight}.

\begin{table}[!t]
    \centering
    \caption{Per-sample inference complexity at $N_{\mathrm{sc}}=48$, $N_{\mathrm{r}}=256$, $M=64$.}
    \label{tab:complexity}
    \begin{tabular}{lcc}
        \hline
        \textbf{Method} & \textbf{Score forwards} & \textbf{FLOPs} \\
        \hline
        LMMSE~\cite{edfors1996ofdm}     & --   & $6.3\times10^{6}$ \\
        CNN~\cite{dong2019deepcnnchannelestimation}  & --   & $1.5\times10^{10}$ \\
        DPS~\cite{kim2026dps} (2000 steps)        & 2000 & $3.3\times10^{14}$ \\
        STMP~\cite{cai2025score} (100 iterations)    & 200  & $3.3\times10^{13}$ \\
        L-STMP ($L = 5$)               & 5    & $8.2\times10^{11}$ \\
        \hline
    \end{tabular}
\end{table}

\section{Conclusion}
\label{sec:conclusion}

This paper proposed a learned score-based STMP (L-STMP) framework for uplink HBF channel recovery under spatial compression. By unfolding turbo iterations into a trainable network, L-STMP replaces the ideal right-rotational invariance assumption of STMP with learned calibration factors that adapt to the actual compression structure. Simulation results showed that L-STMP outperforms linear MMSE, CNN, DPS, and STMP baselines while requiring only $L=5$ unfolded layers --- a $40\times$ inference complexity reduction as compared to the original STMP. Moreover, L-STMP remains robust under both random DFT and PMI-assisted DFT beam selection, whereas  STMP degrades under the latter due to violated right-rotational invariance, confirming that the proposed approach imposes substantially relaxed requirements on the compression matrix.

\bibliographystyle{IEEEtran}
\bibliography{references}

\end{document}